# The Density Dependence of Edge-Localized-Mode Suppression and Pump-out by Resonant Magnetic Perturbations in the DIII-D Tokamak

Q.M. Hu[1], R. Nazikian[1], B. Grierson[1], N.C. Logan[1], J-K. Park[1], C. Paz-Soldan[2], and Q. Yu[3]

[1] Princeton Plasma Physics Laboratory, Princeton NJ 08543-0451, USA
[2] General Atomics, PO Box 85608, San Diego, California 92186-5608, USA
[3] Max-Plank-Institut fur Plasmaphysik, 85748 Garching, Germany

The density dependence of edge-localized-mode (ELM) suppression and density pump-out (density reduction) by $n = 2$ resonant magnetic perturbations (RMPs) is consistent with the effects of narrow well-separated magnetic islands at the top and bottom of the H-mode pedestal in DIII-D low-collisionality plasmas. Nonlinear two-fluid MHD simulations for DIII-D ITER Similar Shape (ISS) discharges show that, at low collisionality ($v^*_e < 0.5$), low pedestal density is required for resonant field penetration at the pedestal top ($n_{e,ped} \approx 2.5 \times 10^{19}$ m$^{-3}$ at $\psi_N \approx 0.93$), consistent with the ubiquitous low density requirement for ELM suppression in these DIII-D plasmas. The simulations predict a drop in the pedestal pressure due to parallel transport across these narrow width ($\Delta\psi_N \approx 0.02$) magnetic islands at the top of the pedestal that is stabilizing to Peeling-Ballooning-Modes (PBMs), and comparable to the pedestal pressure reduction observed in experiment at the onset of ELM suppression. The simulations predict density pump-out at experimentally relevant levels ($\Delta n_e/n_e \approx -20\%$) at low pedestal collisionality ($v^*_e \approx 0.1$) due to very narrow ($\Delta\psi_N \approx 0.01\text{-}0.02$) RMP driven magnetic islands at the pedestal foot at $\psi_N \approx 0.99$. The simulations show decreasing pump-out with increasing density, consistent with experiment, resulting from the inverse dependence of parallel particle transport on collisionality at the foot of the pedestal. The robust screening of resonant fields is predicted between the top and bottom of the pedestal during density pump-out and ELM suppression, consistent with the preservation of strong temperature gradients in the edge transport barrier as seen in experiment.

The control of Edge-Localized-Modes (ELMs) in ITER is essential for minimizing damage to plasma facing components and achieving the ITER scientific mission.[1] The primary method proposed to mitigate or suppress ELMs is by the use of edge Resonant Magnetic Perturbations (RMPs).[1] This method was first demonstrated in DIII-D[2] and thereafter observed on several tokamaks worldwide.[3-9] Various models have been proposed to explain the causes of ELM suppression,[10-13] however important characteristics of ELM suppression remain unexplained. One characteristic of ELM suppression in DIII-D low collisionality ($v^*_e<0.5$) ITER-similar-shape (ISS) plasmas is the low pedestal density $n_{e,ped} \approx 2\text{-}3\times10^{19}$ m$^{-3}$ required for access to ELM suppression.[14,15] This requirement at low collisionality raises the question whether ELM suppression can be achieved for the much higher densities expected in ITER. A further characteristic of ISS plasmas is that RMPs can produce a strong density reduction (called pump-out)[6,16] that rapidly diminishes as the density increases.[2,6] As with ELM suppression, the density dependence of pump-out remains unexplained. Understanding these ubiquitous phenomena in low-collisionality ISS plasmas is critically important for predicting the accessibility of ELM suppression and the level of pump-out in ITER.

In this Letter we show that the density threshold for ELM suppression, the pedestal pressure reduction at the onset of ELM suppression, and the density dependence of pump-out by RMPs in low collisionality DIII-D plasmas, can be understood in terms of (i) the penetration of narrow well-separated magnetic islands at the top and bottom of the H-mode pedestal, and (ii) the robust screening of resonant fields in the gradient region of the Edge-Transport-Barrier (ETB). These conclusions are drawn from nonlinear two-fluid MHD simulations using the TM1[17] code using experimentally relevant profiles, transport coefficients, resistivity and RMP amplitudes in the DIII-D tokamak.

We use ITER-similar-shape (ISS) plasmas[18,19] with the following parameters: toroidal field $B_T = -1.94$ T, normalized beta $\beta_N = 2.0$, pedestal temperature $T_{e,ped} \approx 1$ keV, density $n_{e,ped} \approx 2\text{-}4\times10^{19}$ m$^{-3}$, $Z_{eff} \approx 2.5$, collisionality $v^*_e \approx 0.1\text{-}0.5$ at the top of the pedestal, triangularity $\delta = 0.55$, current $I_p = 1.37$ MA, safety factor $q_{95} = 4.1$, co-$I_p$ neutral beam power $\approx 6$ MW and $\approx 1$ MW central electron cyclotron heating. The DIII-D in-vessel coils (I-coils) are configured to generate magnetic perturbation with toroidal mode number of $n = 2$. The $n = 2$ field in the upper row of coils is rotated toroidally at 1 Hz while the $n = 2$ field in the lower row is held fixed with an I-coil current of 4 kA, producing a sinusoidal resonant ($m/n = 8/2$) magnetic field perturbation on the plasma surface $\psi_N = 1$ as shown in Fig. 1a. Fig.1a shows the relative phase between the upper and lower I-coils (black) and the amplitude of the $m/n = 8/2$ magnetic perturbation at the plasma boundary $\psi_N = 1$, calculated using the ideal MHD code GPEC.[20]

Only the lower density discharge (Fig. 1 blue curves) transitions to ELM suppression at the peak of the resonant field strength (yellow bands), concomitant with a co-$I_p$ increase in the carbon VI toroidal velocity (Fig. 1d) and a reduction in the pedestal electron pressure (Fig. 1e). This transition coincides with a sudden increase in the poloidal field strength (see Fig.2d in ref. 18), and a decrease in the pedestal temperature (see Fig.7 in ref. 21). Experimentally, the upper pedestal density for ELM suppression is around $2\text{-}3\times10^{19}$ m$^{-3}$ in these low collisionality discharges which lies between the minimum densities of the two discharges in Fig.1c. The pedestal density and pressure are obtained from hyperbolic tangent fits to Thomson scattering data.[22]

The higher density discharge in Fig. 1 (#159326, red) with similar $n = 2$ RMP exhibits density pump-out but no ELM suppression. The initial profiles for the two discharges at the minimum of the RMP amplitude are shown in Fig. 1f-i at $t = 3.25$ s for #158115 and $t = 3.4$ s for #159326, including the E×B frequency ($\omega_E = E_r/|RB_\theta|$) profile from charge exchange measurements.

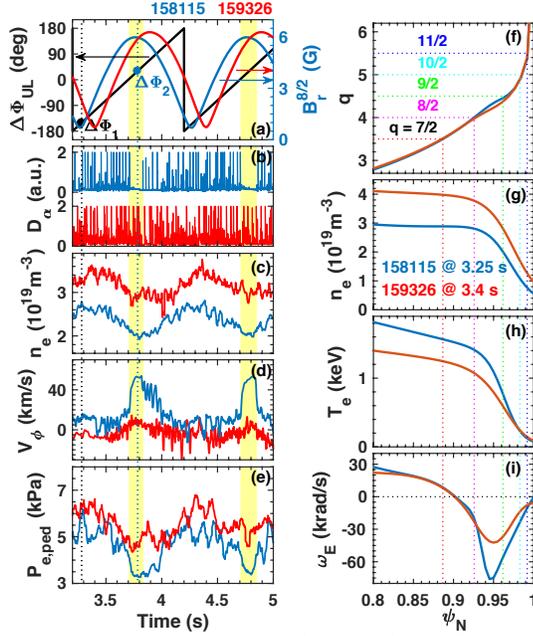

**Fig. 1.** $n = 2$ ELM control experiments with (158115, blue, $P_{NBI} = 6$ MW, $P_{ECH} = 1$ MW) and without (159326, red, $P_{NBI} = 6$ MW, $P_{ECH} = 0$ MW) ELM suppression: (a) relative phase between upper and lower coils $\Delta\Phi_{UL}$ (black), RMP strength for $m/n = 8/2$ calculated by GPEC, (b) $D_\alpha$ near the inner strike, (c) pedestal density $n_{e,\,ped}$, (d) edge impurity velocity $V_\phi$ at $\psi_N = 0.95$ in the co-$I_p$ direction, (e) pedestal electron pressure $P_{e,ped}$. Profiles at the minimum of the RMP: (f) safety factor $q$, (g) electron density $n_e$, (h) temperature $T_e$, and (i) E×B frequency $\omega_E = E_r/|RB_\theta|$. The central line-averaged density is $3.8 \times 10^{19}$ m$^{-3}$ and $4.5 \times 10^{19}$ m$^{-3}$ for #158115 and #159326 when the applied RMP is minimum, respectively.

To model the pedestal response to $n = 2$ RMPs we use the time-dependent resistive cylindrical nonlinear two-fluid MHD model TM1.[17,23,24] TM1 utilizes helical field boundary conditions obtained from full geometry ideal MHD plasma response calculations using GPEC.[20] The TM1 model solves for the balance between the electromagnetic (EM) torque and the plasma viscosity, which governs the threshold for resonant field penetration, and also solves for the enhanced transport produced by resonant field penetration. Previously TM1 studies explored resonant field penetration in the core of TEXTOR,[25] DIII-D[26] and J-TEXT[27] tokamaks in the context of large-scale ($m/n = 2/1$) tearing and locked modes. A description of the TM1 code and coupling with GPEC is provided in the supplemental material.

Multiple helicity magnetic perturbations ($m/n = 7/2$, 8/2, 9/2, 10/2 and 11/2) are included in the TM1 modeling of the pedestal response to RMPs in DIII-D. These helicities correspond to major low order rational surfaces in the pedestal region. These GPEC calculated helical fields at the boundary of TM1 provide the tearing drive for island formation in the pedestal region. The location of these rational surfaces is indicated in Fig. 1f. GPEC[20] is used to evaluate the total (vacuum + plasma response) field for each helical harmonic at the simulation boundary. TM1 takes as input the GPEC fields and initial kinetic equilibrium profiles from experiment when the RMP is negligible (Fig.1f-i). The TRANSP code[28] is used to obtain the momentum diffusivity $\chi_\varphi$ and electron heat conductivity $\chi_e$ from power and momentum balance. The particle diffusivity $D_\perp$ is obtained from TRANSP using the Porter method.[29] These transport coefficients are similar at the top and foot of the pedestal ($D_\perp \approx \chi_\varphi \approx \chi_e \approx 1$ m$^2$/s) for all the discharges studied in this Letter and we use these values throughout. The calculated neoclassical resistivity[30] is used in the simulations, which spans the range $1\times10^{-7}$-$6\times10^{-7}$ $\Omega$m at the pedestal top and $1\times10^{-6}$-$6\times10^{-6}$ $\Omega$m at the foot of the pedestal for the discharges considered in this Letter.

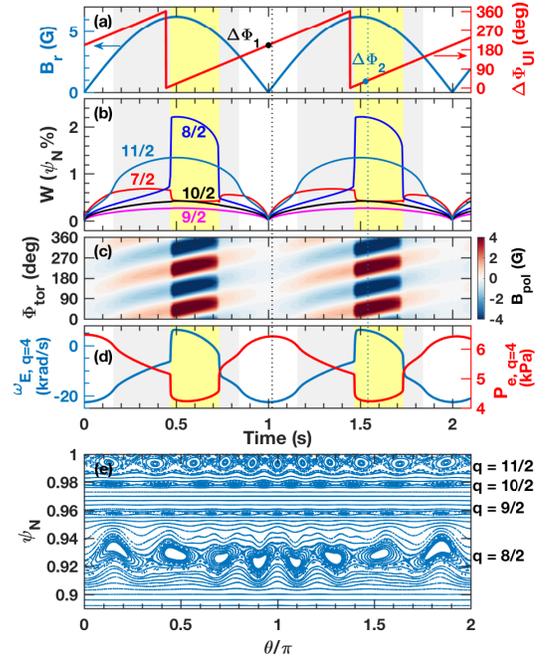

**Fig. 2.** (a) The 8/2 RMP amplitude $B_r$ and phase $\Delta\Phi_{UL}$ at $\psi_N \approx 1$ from GPEC. TM1 output: (b) magnetic island width for $m/n = 7/2$, 8/2, 9/2, 10/2 and 11/2, (c) contour of $n = 2$ $B_{pol}$ (G) versus time and toroidal angle at $\psi_N = 1.1$, (d) $\omega_E$ and electron pressure $P_e$ at the $q = 4$ rational surface, (e) Poincaré plot of the poloidal flux surfaces during 8/2 RMP penetration (at $\Delta\Phi_2$ in Fig. 1 & Fig. 2a).

Nonlinear TM1 simulations for the ELM suppressed discharge (#158115) are shown in Fig.2. The peak helical fields (in Gauss) from GPEC for the total resonant field at the simulation boundary are: $B_r^{11/2} = 16$, $B_r^{10/2} = 11$, $B_r^{9/2} = 8$, $B_r^{8/2} = 6$, $B_r^{7/2} = 2$. For simplicity we only display the $m/n = 8/2$ component in Fig. 2a as the others are in phase. TM1 yield magnetic island widths in $\psi_N$ at the rational surfaces as shown in Fig.2b, with island size < 2% of poloidal flux. The dominant islands are well separated. The modeling shows two dominant magnetic islands, the $m/n = 11/2$ and 8/2 islands at the bottom and top of the pedestal, respectively. From Fig.2b the $m/n = 11/2$ island at the foot of the pedestal (at $\psi_N \approx 0.99$) penetrates at low RMP amplitude ($\delta B/B \approx 5\times10^{-5}$) and exhibits a smooth variation with the applied field. In contrast, the $m/n = 8/2$ island penetrates at the top of the pedestal near the peak of the resonant amplitude. Penetration is characterized by the sudden increase in the $m/n = 8/2$ island width (yellow band in Fig. 2b), poloidal field strength $B_{pol}$ at $\psi_N = 1.1$ (Fig.2c) and $\omega_E$ (blue) at the $q = 4$ surface (Fig. 2d). TM1 also predicts a reduction in the electron pressure $P_e$ at the $q = 4$ surface due to enhanced parallel thermal and particle transport across the magnetic island (Fig. 2d, red curve). The reduction in the pedestal pressure is of the order observed in experiment (see Fig. 1e). The Poincaré plot of the magnetic surfaces during 11/2 and 8/2 penetration (Fig. 2e) reveals two narrow ($\Delta\psi_N \approx 0.01$-$0.02$) island chains, one at the top ($q = 8/2$ at $\psi_N \approx 0.93$) and one at the bottom ($q = 11/2$ at $\psi_N \approx 0.99$) of the pedestal. Strong screening of resonant fields occurs in the gradient region of the pedestal ($\psi_N = 0.94$-$0.98$), which accounts for the preservation of the ETB during pump-out and ELM suppression.

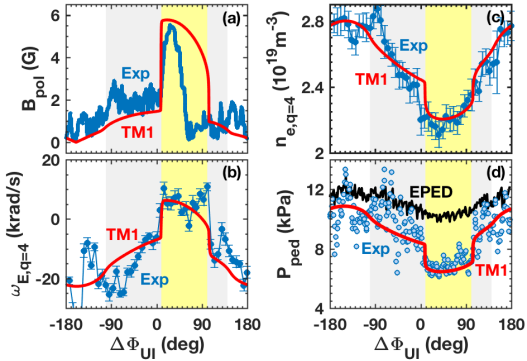

**Fig. 3**. Comparison between experiment (blue) and simulation (red) for: (a) $n = 2$ $B_{pol}$, (b) $\omega_E$, (c) $n_e$ at $q = 4$ rational surface, (d) Total pedestal pressure $P_{ped}$ from experiment (blue circles), EPED prediction of the electron pressure $P_{EPED}$, and TM1 simulation (red curve). Here, the total pedestal pressure is two times of electron pedestal pressure for experiment and TM1 simulation results.

Fig.3 compares the evolution of the plasma pedestal response from TM1 (red) and experiment (blue) for the ELM suppressed discharge (#158115). The simulated $B_{pol}$ at $\psi_N = 1.1$ shows a jump of the same magnitude as in experiment at the onset of ELM suppression. The magnitude of the jump in $\omega_E$, $n_e$ and $P_e$ at the $q = 4$ surface from TM1 is consistent with the change in experiment during ELM suppression (Fig.3b-d). Fig. 3d compares the TM1 prediction of the pedestal pressure (red), the measured pedestal pressure (blue) and the EPED[12] model prediction of the pedestal pressure. Here, the ion pressure is assumed to be the same as electron pressure, as a result, the total pedestal pressure is two times of electron pedestal pressure for experiment and TM1 simulation results. TM1 predicts the pedestal pressure to drop well below the EPED prediction during ELM suppression. The pressure drop arises from enhanced parallel transport across the 8/2 island at the $q = 4$ surface. The EPED prediction of the Peeling-Ballooning-Mode (PBM) threshold is well above the TM1 pedestal pressure prediction during ELM suppression, consistent with the linear stabilization of the PBMs by magnetic island formation at the onset of ELM suppression.[6] This result demonstrates that at the threshold of resonant field penetration in DIII-D, the island size is sufficient to stabilize ELMs.

Significant density pump-out is observed in the simulations correlated with $m/n = 11/2$ resonant field penetration at the foot of the pedestal. The 11/2 field penetrates at much lower amplitude than the 8/2 field and so produces continuous pump-out, in contrast to the sudden onset of ELM suppression with increasing field strength. From the simulations, the magnitude of the pump-out follows the $m/n = 11/2$ island width (Fig.2b gray shaded region). A further increase in pump-out occurs at the onset of $m/n = 8/2$ field penetration, as seen in experiment.

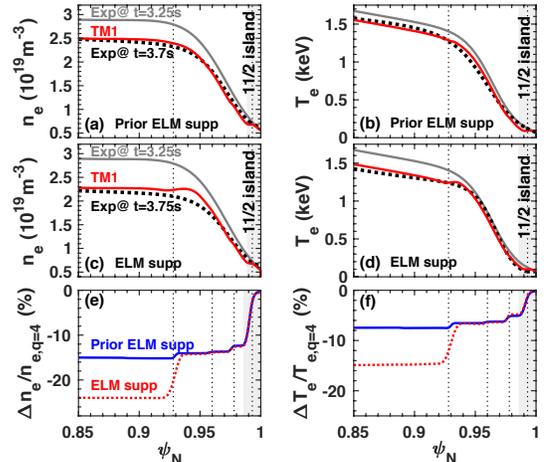

**Fig. 4**. Experimental (black dotted) and TM1 modeling (red solid) for the profiles of $n_e$ and $T_e$ (a, b) prior to ELM suppression ($t = 3.7$ s) and (c, d) during ELM suppression ($t = 3.75$ s) for shot 158115, the initial profiles at $t = 3.25$ s are shown in gray. The simulated relative change in (e) density $\Delta n_e/n_e$ and (f) temperature $\Delta T_e/T_e$ prior to (blue) and during (red) ELM suppression for #158115.

Fig. 4 compares the profiles of $n_e$ and $T_e$ from experiments and simulation for #158115 just before and during ELM suppression. The initial profiles are shown in solid gray lines in Fig.4a-d. The experimental profiles are also shown during the maximum of the applied RMP (black dashed lines), during ELM suppression (Fig.4c,d) and just before ELM suppression (Fig.4a,b). The TM1 simulations are in red. The magnitude of the TM1 predicted changes are similar to experiment just before and during ELM suppression. Fig.4e,f shows the density and temperature change versus radius, normalized to the top of pedestal values. Before suppression most of the density decrease comes from profile flattening across the 11/2 surface, in addition to slight pump-out from the $q$ = 10/2, 9/2 and 8/2 surfaces. Generally, the 11/2 RMP is more effective in decreasing top of pedestal density $n_e$ than $T_e$. This is consistent with TM1 simulation and is due to the density scale length $L_n$ being shorter than $L_{T_e}$ at the pedestal foot near the 11/2 surface.

The trend of pump-out versus density is shown in Fig. 5a,b for a number of low collisionality ISS plasmas. Fig. 5a shows a comparison of the density pump-out ($\Delta n_e/n_e$) versus density at the pedestal top for both experiment (blue) and TM1 simulations (red). The level of pump-out in experiment strongly decreases with increasing density, in close agreement with TM1 prediction. The decrease in the level of pump-out is consistent with the increase in the collisionality at the foot of the pedestal, which varies over the range 0.2-2 for these plasmas. Semi-collisional two-fluid theory[31] and simulation[17,32] reveal that the parallel current perturbation contributes to particle transport through the continuity equation, and this contribution varies inversely with the collisionality. The higher the collisionality the lower the flattening of the density across the 11/2 island and the lower the pump-out.

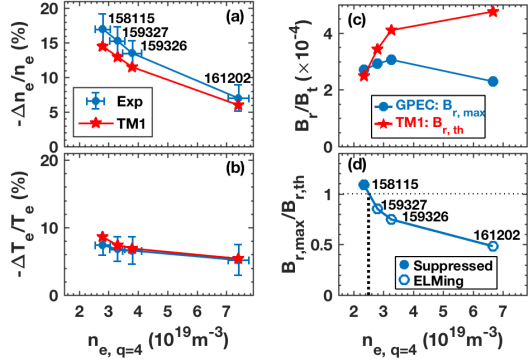

Fig. 5. Comparison of (a) the density pump-out magnitude $-\Delta n_e/n_e$ and (b) $-\Delta T_e/T_e$ versus density at the $q$ = 4 surface for experiment (blue) and TM1 simulations (red), (c) The maximum RMP amplitude at the $m/n$ = 8/2 surface from GPEC ($B_{r,max}$, blue) and the penetration threshold from TM1 ($B_{r,th}$, red), and (d) their ratio $B_{r,max}/B_{r,th}$ versus the density at the $q$ = 4 surface. Discharges that are not ELM suppressed are open blue circles in Fig. 5d.

It is interesting that the peak temperature gradient in the ETB is retained after resonant field penetration in TM1 (Fig. 4d), consistent with experiment. TM1 predicts strong temperature gradients in the ETB due to the effective screening of resonant fields in the gradient region of the pedestal (0.94<$\psi_N$<0.98 in Fig. 2e). Previously, strong pump-out has been identified in simulations resulting from a wide radial region of magnetic stochasticity encompassing the entire ETB.[33] However at low-collisionality and resistivity, strong temperature gradients cannot be maintained in a stochastic magnetic field, as revealed both from experimental measurements[34] and general theoretical[35] considerations. The TM1 simulations for low-collisionality DIII-D ISS plasmas are free of large stochastic regions, consistent with the preservation of the ETB and also close to linear M3D-C1 calculations of effective resonant field screening in similar plasmas.[36]

Fig.5c,d shows the threshold for top-of-pedestal resonant field penetration versus density. In Fig. 5c, the GPEC calculated $m/n$ = 8/2 RMP amplitude $B_r$ (blue) and the TM1 calculated penetration threshold $B_{r,th}$ (red) are shown versus pedestal density. TM1 predicts an increase in the $m/n$ = 8/2 penetration threshold with increasing density. The threshold exceeds the available RMP amplitude for $n_{e,ped}$>2.5×10$^{19}$ m$^{-3}$ (black dashed line Fig. 5d), consistent with the ubiquitous density threshold for ELM suppression in DIII-D.[14,15] The three high density discharges with $B_r/B_{r,th}$ < 1 are not ELM suppressed and have no observable field penetration event at the pedestal top. The increasing penetration threshold $\delta B_{r,th}/B_T$ with increasing density is due to the increasing inertia of the plasma when the flow frequency is constant, requiring higher RMP levels that eventually exceed the experimentally available RMP amplitude for penetration.

We should note that the TM1 code does not include the non-resonant plasma kink response. While there has been consideration that the kink response can drive pump-out[37] and excite ballooning modes,[38,39] recent global gyro-kinetic simulations with the GTC[40] and XGC[41] codes do not show significant contribution of the kink response on finite-n ballooning stability and neoclassical cross-field transport for the low-collisionality plasmas studied in this Letter. This leaves the tearing response for which TM1 is well suited for performing analysis at ITER relevant resistivity and collisionality.

Here we do not claim that other mechanisms are irrelevant for ELM suppression. The low-n plasma kink response may have transport and stability effects in conditions that differ substantially from DIII-D ISS plasmas. Also, there is evidence that the nonlinear interaction of RMPs with PBMs[10,11] can account for ELM mitigation,[42,43] and may be key to understanding ELM suppression at high pedestal collisionality.[2,5] Finally, enhanced micro-turbulence[44] is correlated with the penetration of resonant fields[18,45] and these could act synergistically with magnetic islands to suppress ELMs.

In this Letter we have addressed the key phenomenology of ELM suppression and density pump-out in low-collisionality DIII-D ISS plasmas using the nonlinear two-fluid TM1 model. We have shown that:

1) The formation of magnetic islands at the foot of the pedestal produces experimentally relevant density pump-out that decreases with increasing density,

2) Resonant field penetration at the top of pedestal requires low-density for DIII-D ISS plasmas, consistent with the low-density requirement for ELM suppression in low-collisionality DIII-D plasmas,

3) Magnetic island formation at the top of the pedestal enhances particle and thermal transport, resulting in a pedestal pressure reduction that is stabilizing to PBMs,

4) Strong temperature gradients persist during ELM suppression due to the effective screening of resonant fields between the top and bottom of the pedestal.

While the TM1 model combined with GPEC is successful in addressing experimental trends in low-collisionality DIII-D plasmas, more work is required to test the model in other plasma conditions in DIII-D and in other tokamaks worldwide, particularly those operating at ITER-relevant collisionality.

**Supplementary Material**

The details of the TM1 and GPEC models as well as the analysis procedure are introduced in the supplementary material.

**Acknowledgements**

This material is based upon work supported by the U.S. Department of Energy, Office of Science, Office of Fusion Energy Sciences, using the DIII-D National Fusion Facility, a DOE Office of Science user facility, under Awards DE-AC02-09CH11466, DE-FOA-0001386, DESC0015878 and DE-FC02-04ER54698.